\documentclass[fleqn,usenatbib]{mnras}

\usepackage{newtxtext,newtxmath}

\usepackage[T1]{fontenc}
\usepackage{adjustbox}
\usepackage{tablefootnote}

\DeclareRobustCommand{\VAN}[3]{#2}
\let\VANthebibliography\thebibliography
\def\thebibliography{\DeclareRobustCommand{\VAN}[3]{##3}\VANthebibliography}

\usepackage{graphicx}	
\usepackage{adjustbox}
\usepackage{arydshln}
\usepackage{longtable}
\usepackage{lscape}

\defcitealias{Bhattacharya25}{Bh+25}
\defcitealias{MD14}{MD14}

\title[Chemical enrichment sequences of SDSS SFGs out to z$\sim$0.3 from O \& Ar abundances]{Mass-dependent chemical enrichment sequences of SDSS star-forming galaxies out to z$\sim$0.3 revealed by direct O \& Ar abundances}

\author[Souradeep Bhattacharya]{
Souradeep Bhattacharya,$^{1, 2}$\thanks{E-mail: s.bhattacharya3@herts.ac.uk},
Magda Arnaboldi$^{3}$,
Chiaki Kobayashi$^{1}$,
Ortwin Gerhard$^{4}$ and
Kanak Saha$^{2}$
\\
$^{1}$Centre for Astrophysics Research, Department of Physics, Astronomy and Mathematics, University of Hertfordshire, Hatfield AL10 9AB, UK\\
$^{2}$Inter University Centre for Astronomy and Astrophysics, Ganeshkhind, Post Bag 4, Pune 411007, India\\
$^{3}$European Southern Observatory, Karl-Schwarzschild-Str. 2, 85748 Garching, Germany\\
$^{4}$Max-Planck-Institut für extraterrestrische Physik, Giessenbachstraße, 85748 Garching, Germany
}

\date{Accepted XXX. Received YYY; in original form ZZZ}

\pubyear{2025}

\begin{document}
\label{firstpage}
\pagerange{\pageref{firstpage}--\pageref{lastpage}}
\maketitle

\begin{abstract}
Individual stars in the Milky Way (MW) and its satellites have been shown to trace galaxy stellar mass dependent sequences in the $\alpha$-abundance ([$\alpha$/Fe]) vs metallicity ([Fe/H]) plane. Testing the universality of such sequences has been elusive as deep absorption-line spectra required for [$\alpha$/Fe] and [Fe/H] measurements beyond the local group are mostly limited to integrated light from nearby, relatively high-mass, early-type galaxies. However, analogous to [$\alpha$/Fe] vs [Fe/H] for stars, we now have  log(O/Ar) vs 12+log(Ar/H) for the integrated nebular light of star-forming galaxies (SFGs). From Sloan-Digital Sky Survey (SDSS) observations of $\sim3000$ SFGs out to z$\sim0.3$, where we directly determined O \& Ar abundances, we obtain for the first time the distribution of an ensemble of SFGs in the log(O/Ar) vs 12+log(Ar/H) plane. We show that higher (<M$\rm_{*}$>$\sim 2.6 \times 10^9$M$_{\odot}$) and lower mass (<M$\rm_{*}$>$\sim 1.7 \times 10^7$M$_{\odot}$) SFGs clearly trace distinct mass dependent sequences in this plane. Such sequences are consistent with expectations from galaxy chemical evolution (GCE) models of SFGs that are driven primarily by the interplay of core-collapse and Type Ia supernovae. 
\end{abstract}

\begin{keywords}
galaxies: abundances -- supernovae: general -- galaxies: formation -- galaxies: evolution -- Galaxy: evolution
\end{keywords}




\section{Introduction} 
\label{sec:intro}

The locus traced by Milky Way (MW) stars in the  $\alpha$-abundance ([$\alpha$/Fe]) vs. metallicity ([Fe/H]) diagram has, for a long time, informed our understanding of galaxy chemical enrichment \citep[][]{Tinsley79}. Stars formed from interstellar medium (ISM) only enriched by core-collapse supernovae (CCSNe) exhibit the highest [$\alpha$/Fe] values \citep[e.g.][]{Pagel97,kob20sr}. Subsequent generations of stars formed from ISM after Type Ia supernovae (SNe Ia) explosions, that released more Fe into the ISM than previously, showcase a decreasing trend in the [$\alpha$/Fe] vs [Fe/H] diagram \citep[e.g.][]{Matteucci86,Edvardsson93, Fuhrmann98}. 

Such a trend, with a plateau in [$\alpha$/Fe] at low [Fe/H] followed by a linearly decreasing trend in [$\alpha$/Fe] with increasing [Fe/H], is characteristic of a self-regulated (see also Section~\ref{sec:mzr}) chemical enrichment scenario that may be described with a galaxy chemical enrichment (GCE) model having enrichment from star-formation balancing gas inflow/outflows  \citep[e.g.][]{kob20sr}. In such a scenario, the star-formation efficiency of a galaxy stellar population, characterized by it mass and star-formation history, sets the locus of its stars in the [$\alpha$/Fe] vs [Fe/H] diagram.

MW thick disc and bulge stars exhibit higher [$\alpha$/Fe] at given [Fe/H] than thin disc stars \citep[e.g.][]{Bensby14,hayden15}. This is a consequence of the former having relatively higher star-formation efficiency at early times compared to the latter. Stars in the Magellanic clouds and other dwarf MW satellite galaxies, having even lower star-formation efficiencies than the MW thin disc, showcase even lower [$\alpha$/Fe] values at given [Fe/H] \citep{Pompeia08,Tolstoy09,Kirby11}. The dwarf galaxy stars thus trace decreasing loci in the [$\alpha$/Fe] vs [Fe/H] plane, nearly-parallel to that of the MW regions, but at lower and lower [$\alpha$/Fe] with decreasing galaxy mass (see Figure~6 in \citealt{KobayashiPhilip23}). 

Beyond the local group, [$\alpha$/Fe] and [Fe/H] determination from deep absorption-line spectra of faint individual stars are limited with current instrumentation. However, such measurements become possible from integrated stellar spectra of early-type galaxies \citep[ETGs; e.g.][]{Trager00,Thomas05, Iodice19}. When the [$\alpha$/Fe] and [Fe/H] are measured for an individual ETG, a median estimate of these properties is obtained for its entire stellar population. Using velocity dispersion ($\rm\sigma$) as a proxy for galaxy mass, \citet[][]{Sybilska18} show that at fixed [Fe/H], ETGs with lower $\rm\sigma$ show lower [Mg/Fe]  (in lieu of [$\alpha$/Fe]) values. 
[$\alpha$/Fe] and [Fe/H] determination from integrated absorption-line spectra confines analysis of chemical enrichment sequences to nearest relatively higher mass ETGs ($>10^{9}$~M$_{\odot}$; \citealt{Sybilska18}). \citet{Zahid17} utilized stellar population synthesis models fitted to stacked spectra of star-forming galaxies (SFGs) to determine their [Fe/H] but such a technique is not possible for fainter individual SFGs. To overcome observational limitations and test the effect of mass dependence on the chemical enrichment of a large population of individual galaxies, we require determination of equivalent quantities to [$\alpha$/Fe] and [Fe/H] but for SFGs, that constitute the vast majority of galaxies in the universe. 

\citet{Arnaboldi22} showed that the log(O/Ar) vs 12 + log(Ar/H) plane for emission nebulae\footnote{Planetary nebulae \citep{Bh+19, Bh+19b,Bh21,Bhattacharya22,Bhattacharya23a} and HII regions \citep{Esteban20} having direct O \& Ar abundances surveyed in the disc of M31. We found high and low $\rm\alpha$ stellar populations in the M31 thick and thin disc respectively, with the latter formed following gas infall $\sim$2--4~Gyr ago \citep[][]{Arnaboldi22,Kobayashi23}.
} is analogous to the [$\alpha$/Fe] vs [Fe/H] plane for stars. \citet{Kobayashi23}  showed that with well-constrained GCE models, the log(O/Ar) vs 12 + log(Ar/H) plane can be connected to the [$\alpha$/Fe]-[Fe/H] plane. Like Fe, SNe Ia preferentially produce more Ar than light $\alpha$-elements like O, whereas CCSNe produce near-constant log(O/Ar), see \citet[][]{kob20sr}. 

Through detection of the temperature sensitive [OIII]$\lambda$ 4363 \AA~auroral line, determination of O \& Ar abundances has been carried out for SFGs out to z$\sim$7.7 \citep[e.g.][]{Izotov06,Cordova24,Stanton24,Bhattacharya25, Bhattacharya26}.  Utilisation of the log(O/Ar) vs 12 + log(Ar/H) plane for SFGs, and subsequent interpretation of galaxy chemical enrichment at high-redshifts (z$\sim$1.3--7.7), was first demonstrated in \citet{Bhattacharya25}.


The emission-line spectrum of each SFG reflects its constituent HII regions and thus its determined O \& Ar abundances are associated with the ISM ionised by its youngest stars. The abundances thus reflect the cumulative chemical enrichment of an SFG by its previous generations of stars. Individual SFGs that underwent different star-formation histories will have ISM with different states of chemical enrichment, thus occupying different positions in the log(O/Ar) vs 12 + log(Ar/H) plane.  An ensemble of SFGs, if governed by the same chemical enrichment mechanisms, should trace out any underlying trend in the log(O/Ar) vs 12 + log(Ar/H) plane. Analogous to stars within a galaxy that showcase any imprints of its governing chemical enrichment mechanisms and star-formation efficiency in the [$\alpha$/Fe] vs [Fe/H] plane, individual SFGs may also showcase such imprints on their ensemble in the log(O/Ar) vs 12 + log(Ar/H) plane.  By determining the positions in this plane of individual SFGs covering a range of stellar masses, we can examine whether stellar mass dependence of galaxy chemical enrichment mechanisms, as observed in the Local Group from the stars in the MW and its dwarf satellites, is also prevalent in ensembles of SFGs.

In this work, we present the chemical enrichment sequences traced by SFGs of different stellar masses out to z$\sim0.3$ from Sloan Digital Sky Survey (SDSS) spectroscopy. The sample selection of SFGs from catalogue data and direct determination of their O \& Ar abundances, as well as further pruning of the sample from the mass-metallicity relation, is presented in Section~\ref{sec:sample}. The positions of these galaxies in the log(O/Ar) vs 12 + log(Ar/H) plane and the sequences traced by SFGs of different stellar masses is presented in Section~\ref{sec:analysis}. The implications are discussed and concluding remarks are made in Section~\ref{sec:discussion}.


\section{Abundance determination and sample selection} 
\label{sec:sample}


\subsection{SDSS Catalogue Data} 
\label{sec:data}

\begin{figure}
\centering
\includegraphics[width=\columnwidth]{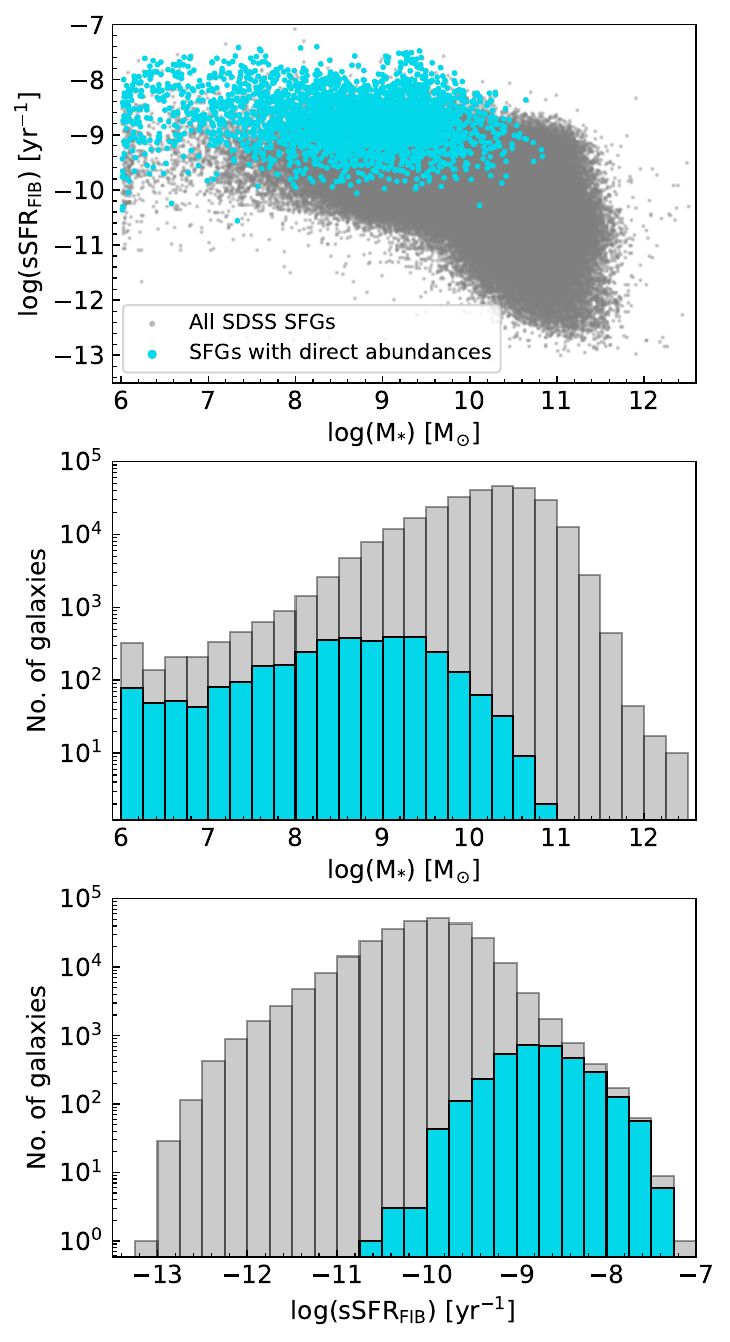}
\caption{[Top] Distribution of logarithm of total stellar mass vs. log(sSFR$\rm_{FIB}$) of all SFGs out to z$\sim$0.3 from the MPA-JHU SDSS catalogue (grey; see also Section~\ref{sec:data}) and the 3306 SFGs with direct O and Ar abundances determined (cyan; see also Section~\ref{sec:abund}). [Middle] Histogram of logarithm of total stellar mass of all SFGs and those with direct O and Ar abundances. Note that number of galaxies are shown in log scale. [Bottom] Same as Middle but for log(sSFR$\rm_{FIB}$) of the same SFG samples.} 
\label{Fig:sample}
\end{figure}

We utilize the tabulated emission-line  fluxes ([OII]$\lambda\lambda$ 3726,3729 \AA, H$\delta$, H$\gamma$, [OIII]$\lambda$ 4363 \AA, H$\beta$, [OIII]$\lambda\lambda$ 4959,5007 \AA, H$\alpha$, [SII]$\lambda\lambda$ 6717,6731 \AA~\& [ArIII]$\lambda$ 7136 \AA) and derived properties (redshift: z; total stellar mass: M$\rm_{*}$; fibre specific star-formation rate: sSFR$\rm_{FIB}$; SUBCLASS; \citealt{Brinchmann04}) of galaxies from the SDSS DR8 \citep{Aihara11} presented in the \href{https://www.sdss4.org/dr17/spectro/galaxy_mpajhu}{MPA-JHU SDSS catalogue}. The spectra have a wavelength coverage of 3800--9200\AA. As the red-most emission-line of interest required for O and Ar abundance determination ([ArIII]$\lambda$ 7136 \AA) was only tabulated for galaxies out to z=0.2853, we have limited our sample to this redshift. Only galaxies having reliable redshift determinations (Z\_warning=0) have been considered. To restrict our analysis to only SFGs, we limit our sample to galaxies having SUBCLASS =`STARBURST' or `STARFORMING'. To enable the subsequent analysis, only those SFGs having derived stellar mass (M$\rm_{*}$) and specific star-formation rate (sSFR$\rm_{FIB}$\footnote{We discuss the need and impact of utilizing sSFR$\rm_{FIB}$ instead of sSFR$\rm_{TOT}$ for this work in Appendix~\ref{app:fib_tot}.}) values in the catalogue have been considered in the sample. This sample of SFGs out to z$\sim$0.3 thereby consists of 280902 galaxies and is noted in Table~\ref{tab:sample}. Their log(M$\rm_{*}$) vs. log(sSFR$\rm_{FIB}$) distribution as well as log(M$\rm_{*}$) and log(sSFR$\rm_{FIB}$) histograms are shown in Figure~\ref{Fig:sample}.


\begin{table}
\caption{Sample sizes of SDSS SFGs out to z$\sim$0.3 selected from the MPA-JHU SDSS catalogue and described in this work (see Section~\ref{sec:data}).}
\centering
\adjustbox{max width=\columnwidth}{
\begin{tabular}{lr}
\hline
Sample & No. of SFGs \\
\hline 
All SDSS SFGs &  280902 \\
SFGs with direct abundances &  3306\\
Higher mass (M$\rm_{*}>9$~M$\rm_{\odot}$) SFGs with direct abundances & 1256 \\
Lower mass (M$\rm_{*}<8$~M$\rm_{\odot}$) SFGs with direct abundances & 719 \\
SFGs with direct abundances following the empirical MZR & 2370  \\
Higher mass (M$\rm_{*}> 9$~M$\rm_{\odot}$) MZR sub-sample & 879 \\
Lower mass (M$\rm_{*}<     8$~M$\rm_{\odot}$) MZR sub-sample & 503 \\
\hline
\end{tabular}
\label{tab:sample}
}
\end{table}

\subsection{Abundance determination} 
\label{sec:abund}

Amongst these SFGs, only 3716 had both the required faint [OIII]$\lambda$ 4363 \AA~\& [ArIII]$\lambda$ 7136 \AA~ emission lines observed (other lines were either brighter or not essential). For these 3716 SFGs, we then compute their O and Ar abundances using NEAT \citep[Nebular Empirical Analysis Tool;][]{Wesson12}, which applies an empirical scheme to calculate the extinction and elemental abundances. 

For each SFG, NEAT calculates the intrinsic balmer decrement, c(H$\beta$), using the flux-weighted ratios of H$\alpha$/H$\beta$, H$\gamma$/H$\beta$ and H$\delta$/H$\beta$ (whichever pairs are observed) and the extinction law of \citet{Cardelli89}, first assuming a nebular temperature (T$\rm_{e}$) of 10000K and an electron density (n$\rm_{e}$) of 1000 cm$^{-3}$, and then recalculating c(H$\beta$) at the measured T$\rm_{e}$ and n$\rm_{e}$, using an iterative process from the relevant diagnostic lines \citep[see][section 3.3]{Wesson12}. NEAT utilizes the temperature-sensitive [OIII]$\lambda$ 4363 \AA~ line, and the density-sensitive [OII]$\lambda\lambda$ 3726,3729 \AA~ and [SII]$\lambda\lambda$ 6717,6731 \AA~ doublets to obtain T$\rm_{e}$ and n$\rm_{e}$ respectively. For SFGs where we do not observe the required doublets to determine n$\rm_{e}$, 1000 cm$^{-3}$ continues to be assumed as it is expected to have negligible impact on the determined abundances \citep[E.g.][]{Ferland13}. 

NEAT assumes the same T$\rm_{e}$ and n$\rm_{e}$ for all elemental abundance determinations. There are established relations for T$\rm_{e}$ (OIII) vs T$\rm_{e}$ (OII) which allow for distinct assumptions of low and medium ionisation zone temperatures  but these are calibrated on metal-rich HII regions (e.g. \citealt{Pilyugin09}). There is more scatter in T$\rm_{e}$ (OIII) vs T$\rm_{e}$ (OII) at lower metallicities (i.e., at higher Te; e.g. see the compiled low-z lit. values in Fig 7 in \citealt{Cataldi25}). The standard deviation around the 1:1 line in T$\rm_{e}$ (OIII) vs T$\rm_{e}$ (OII) is at most 1200 K (highest scatter we found in the literature, albeit for HII regions in NGC 2403 by \citealt{Rogers21}), which results in an error in O abundance of $\sim$0.1 dex from considering the O++ temperature for the O+ ionic abundance. As this error is small, we use the same T$\rm_{e}$ (OIII) throughout the nebula. Furthermore, Ar$^{3+}$ resides in the same medium ionisation zone as O$^{2+}$ so the Ar abundance remains reliable.

Direct O and Ar ionic abundances are determined from the measured fluxes of the O ([OII]$\lambda\lambda$ 3726,3729 \AA~,  [OIII]$\lambda\lambda$ 4363,4959,5007~\AA) and Ar ([ArIII] $\lambda$ 7136) lines respectively. Ionisation correction factors (ICF) for O is negligible when lines pertaining to both O$^{2+}$ (i.e, [OIII] $\lambda\lambda$ 5007, 4959, 4363 \AA) and  O$^{+}$ ([OII] $\lambda\lambda$ 3727, 3729 \AA) are observed. Elemental Ar abundances are obtained from the Ar$^{2+}$ ionic abundances utilizing the ICF scheme by \citet{Amayo21} that has been found to be suitable for such determinations for HII regions and SFGs \citep[][see also \citealt{Cordova24} and Appendix C in \citealt{Bhattacharya25}]{Esteban25}.

Amongst these 3716 SFGs with O \& Ar abundances determined, we obtain a small number of sources that have c(H$\beta$)=0 or T$\rm_{e}> 35000$ K or n$\rm_{e}> 10000$ cm$^{-3}$. Our abundance determination procedure is not applicable to such sources \citep{Wesson12} which are thus excluded. Our sample of SDSS SFGs out to z$\sim$0.3 with determined O \& Ar abundances thus consists of 3306 galaxies, noted in Table~\ref{tab:sample}.


\begin{figure*}
\centering
\includegraphics[width=\columnwidth]{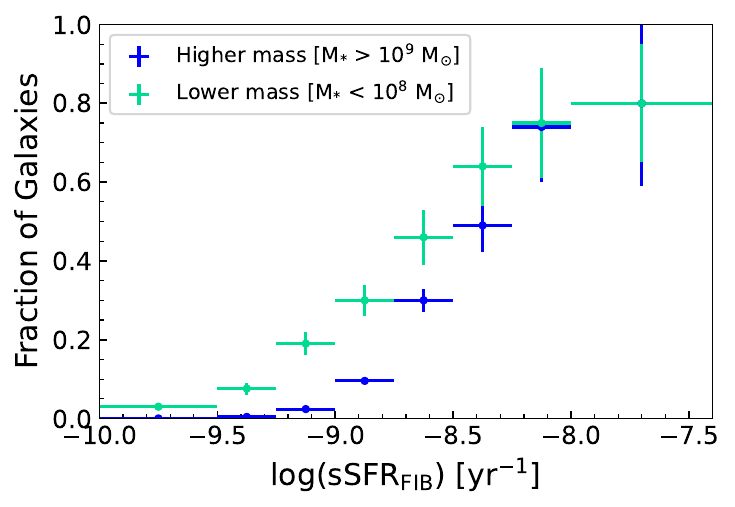}
\includegraphics[width=\columnwidth]{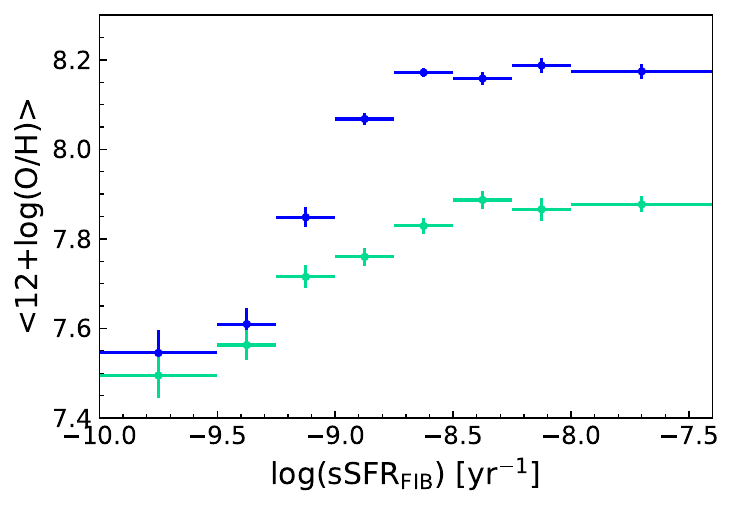}
\caption{[Left] Fraction of SFGs with direct O and Ar abundances determined in different sSFR$\rm_{FIB}$ bins, separately for relatively higher and lower mass SFGs. [Right] The mean O abundances of SFGs in different sSFR$\rm_{FIB}$ bins, separately for relatively higher and lower mass ones. } 
\label{Fig:completeness}
\end{figure*}

\begin{figure*}
\centering
\includegraphics[width=\textwidth]{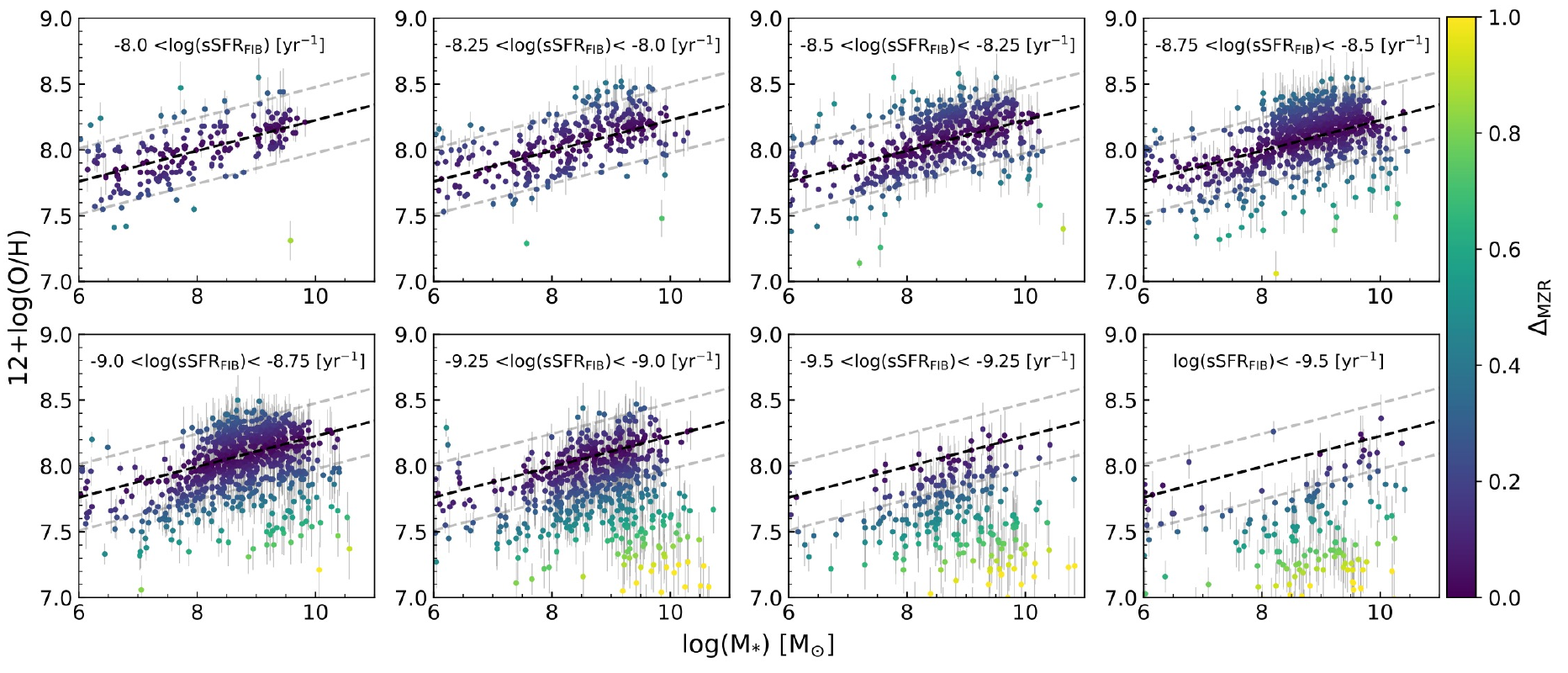}
\caption{log(M$\rm_{*}$) vs 12+log(O/H), shown for SFGs with direct determination of O \& Ar abundances, separated in bins of log(sSFR$\rm_{FIB}$). Error in the determined 12+log(O/H) is marked. The dashed black line is the empirical best-fit for log(M$\rm_{*}$) vs 12+log(O/H) for the highest log(sSFR$\rm_{FIB}$) bin. The SFGs in each panel are coloured by their perpendicular distance from this empirical best-fit MZR, $\rm\Delta_{MZR}$. The dashed grey lines mark the zone where 12+log(O/H) is within 0.25 dex of this line at any given log(M$\rm_{*}$).} 
\label{Fig:mmr}
\end{figure*}

\subsection{Selection completeness effects} 
\label{sec:final}

The log(M$\rm_{*}$) distribution of our sample of SFGs with direct abundances determined is also shown in Figure~\ref{Fig:sample} [Middle]. For relatively lower mass SFGs (M$\rm_{*}<10^8$~M$_{\odot}$), a nearly constant fraction have abundances determined with increasing log(M$\rm_{*}$). However as we move to more massive SFGs, a decreasing fraction of galaxies have determined abundances, with none at M$\rm_{*}>10^{11}$~M$_{\odot}$. Another prominent selection effect can be seen in Figure~\ref{Fig:sample} [Bottom] from the log(sSFR$\rm_{FIB}$) distribution. A high fraction of galaxies have their O \& Ar abundances determined when log(sSFR$\rm_{FIB}$) values are high; but with decreasing log(sSFR$\rm_{FIB}$) values, the fractions of SFGs which have direct abundances determined decreases significantly with none such SFG at log(sSFR$\rm_{FIB}$)$<-10.75$~yr$^{-1}$. The combined effect of the stellar mass and sSFR selection is clear in the log(M$\rm_{*}$) vs. log(sSFR$\rm_{FIB}$) distribution (Figure~\ref{Fig:sample} [Top]), showing that our sample of SDSS SFGs with determined O \& Ar abundances consists mostly of lower mass SFGs (almost all M$\rm_{*}\leq10^{10}
$~M$_{\odot}$) that are undergoing a strong burst of star-formation (almost all log(sSFR$\rm_{FIB}$~$>-10$~yr$^{-1}$).

Figure~\ref{Fig:completeness} [Left] shows the fraction\footnote{Given the small number of SFGs with direct abundances in our sample for each log(sSFR$\rm_{FIB}$) bin, we compute the fraction and uncertainty (95\% confidence interval) using the binomial proportion confidence-interval obtained with the Wilson score interval method \citep{Wilson27}.} of SFGs with direct O \& Ar abundances determined as a function of log(sSFR$\rm_{FIB}$), separately for higher (M$\rm_{*}>10^9$~M$_{\odot}$) and lower mass (M$\rm_{*}<10^8$~M$_{\odot}$) SFGs. The sample sizes are noted in Table~\ref{tab:sample}. The bins of log(sSFR$\rm_{FIB}$) are 0.25 dex wide, except the first and last bins which are wider to accommodate the remaining burstiest and least bursty SFGs respectively in our direct abundance sample. In either case, $\sim80\%$ of SDSS SFGs with log(sSFR$\rm_{FIB}$)$>-8$~yr$^{-1}$have direct abundances determined. As we move to decreasing values of  log(sSFR$\rm_{FIB}$) in Figure~\ref{Fig:completeness} [Left], smaller fractions of SFGs have direct abundances determined. The lowest sSFR bin shown has only 0.03\% and 3\% of higher and lower mass SFGs respectively with direct abundances determined. 

Thus the SFGs in the SDSS sample that are undergoing the strongest bursts, i.e., forming the highest fraction of their stellar mass in the present star-forming episode (hence having highest sSFR), have the brightest emission lines, including the faint [OIII]$\lambda$ 4363 \AA~line that enables direct abundance determination for 80\% of such SFGs.  Meanwhile, for SFGs having decreasing sSFR (i.e., forming a smaller fraction of their stellar mass in the present star-forming episode), their emission lines become increasingly fainter such that the faint [OIII]$\lambda$ 4363 \AA~line that enables direct abundance determination is not detected for a large fraction of SFGs. For decreasing sSFR bins, we illustrate the decreasing star-formation in the SFGs at any given stellar mass by comparing with the objective star-forming main-sequence from \citet{Renzini15} in Appendix~\ref{app:SFMS}.

Figure~\ref{Fig:completeness} [Left] shows that the fraction of higher mass SFGs with directly determined abundances is smaller and decreases more rapidly than that of lower mass SFGs. To characterize such observational selection effect, we first consider the mass-metallicity relation (MZR) of galaxies (discussed further in Section~\ref{sec:mzr}), whereby galaxies on average show increasing O abundances with increasing mass \citep[E.g.][]{Pagel81, Tremonti04, Curti20}. We further consider that the flux of the [OIII]$\lambda$ 4363 \AA~line (for the same [OIII]$\lambda$ 5007 \AA~line flux) is inversely correlated with metallicity, and hence the [OIII]$\lambda$ 4363 \AA~line is detected with higher S/N in relatively metal-poor galaxies \citep{Curti20}. 

Figure~\ref{Fig:completeness} [Right] shows that the mean oxygen abundance of higher mass SFGs with directly determined abundances is clearly higher than that of lower mass ones for the highest sSFR bins (as expected from the MZR). However, as we move to lower log(sSFR$\rm_{FIB}$) bins, the mean oxygen abundances of both the higher and lower mass SFG samples reduce. A possible explanation is that when SFGs are forming a smaller fraction of their stellar mass, i.e., when they have lower sSFR with generally weaker emission lines, the faint [OIII]$\lambda$ 4363 \AA~line (that enables direct abundance determination) remains detected only in the most metal-poor SFGs. 

To consider the selection completeness effects further, we explore the MZR in the same sSFR bins in the subsequent section.

\begin{figure*}
\centering
\includegraphics[width=0.95\textwidth]{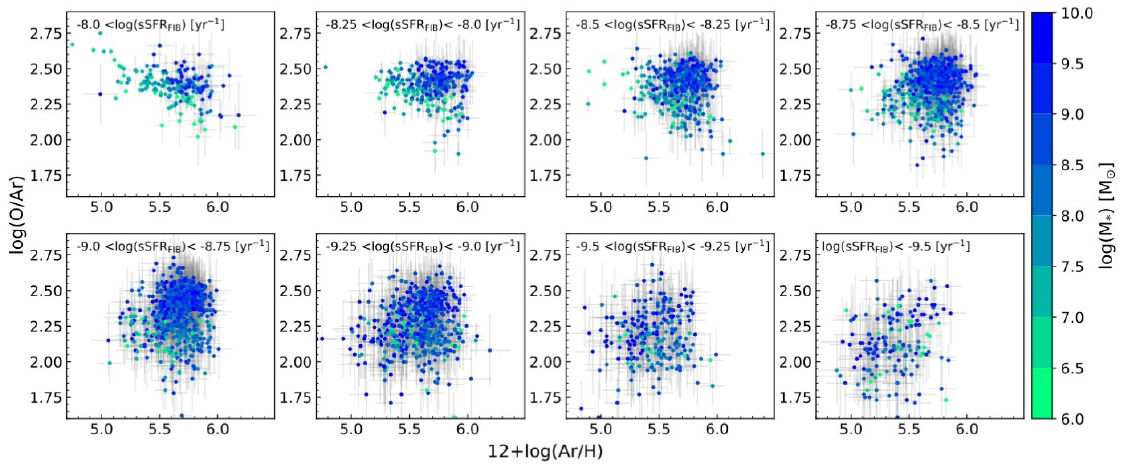}
\caption{Distribution of SFGs in the abundance sample (see Section~\ref{sec:final}) in the log(O/Ar) vs 12+log(Ar/H) plane, colored by their log(M$\rm_{*}$), separated in bins of log(sSFR$\rm_{FIB}$). Error bars are shown in grey.} 
\label{Fig:oar_obs}
\end{figure*}

\subsection{Empirical selection on the mass-metallicity relation (MZR)} 
\label{sec:mzr}

In “self-regulated” galaxies, i.e., a system with gas infall balanced by star formation, the gas phase metallicity approaches the net nucleosynthesis yields (see the analytic equation in \citealt{Tinsley80}), and the difference in the net yields between galaxies of different masses leads to the MZR (see also hydrodynamical simulations in \citealt{Kobayashi07}). In contrast, an extreme gas-infall event or an extremely rapid star burst, will populate galaxies below or above the MZR\footnote{Secondary dependence of the MZR on star-formation rate (or equivalently for sSFR) has been characterised as the fundamental MZR -- FMR \citep{Mannucci10, Curti20}. An extreme gas-infall event would produce SFGs with much lower metallicities than the variation within FMR.} respectively \citep[e.g.][]{Koppen05}.

Figure~\ref{Fig:mmr} shows the  log(M$\rm_{*}$) vs. 12+log(O/H) for the abundance sample SFGs. The top-left panel corresponds to the highest log(sSFR$\rm_{FIB}$) bin, i.e., log(sSFR$\rm_{FIB}$) > -8 yr$^{-1}$ where $\sim80\%$ of SDSS SFGs have direct abundances determined. Within the SDSS survey, such SFGs are forming the largest fraction of their stellar mass in the present star-forming episode (otherwise the sSFR would not be as high), and are thus expected to be self-regulated with the chemical enrichment from past generations of stars imprinted on the measured ISM abundances. These SFGs show a clear linear relationship between galaxy stellar mass and O abundance. 
We characterize this with the best-fit line: $\rm 12+log(O/H)= 0.116 \times log(M_{*}) + 7.06$, that reflects the median O abundance as a function of galaxy mass. $\sim$90\% of SFGs in this bin lie within 0.25~dex of this best-fit line. We utilise this empirical best-fit MZR as a guide to characterise the SFGs in the lower sSFR bins, where our abundance sample is less complete.

We computed the perpendicular distance from this empirical best-fit MZR, $\rm\Delta_{MZR}$, for each SFG with direct abundances determined in our sample. The SFGs in each panel are coloured by their $\rm\Delta_{MZR}$ in Figure~\ref{Fig:mmr}. For the highest log(sSFR$\rm_{FIB}$) bins, since a high fraction of their stellar mass is formed in the present burst of star-formation, most SFGs are self-regulated and thus lie on the empirical MZR having low $\rm\Delta_{MZR}$. With decreasing log(sSFR$\rm_{FIB}$), SFGs form lower and lower fractions of their stellar mass in the present burst of star-formation, and we increasingly find SFGs that are offset from the empirical MZR (thus not self-regulated systems), typically with lower O abundance, having larger $\rm\Delta_{MZR}$ values. Given the preferential detection of the [OIII]$\lambda$ 4363 \AA~line for relatively metal-poor galaxies and the increasing incompleteness of the abundance sample as we move towards lower log(sSFR$\rm_{FIB}$) bins, such metal-poor galaxies become increasingly numerous. These metal-poor SFGs possibly have the most recent burst of star-formation (that gives rise to the observed emission lines) induced by metal-poorer gas infall \citep[e.g.][]{Koppen05}, hence not consistently enriched by the previous generations of stars.  

We can utilize the empirical best-fit MZR to select SFGs that follow the MZR for all log(sSFR$\rm_{FIB}$) bins and thus select SFGs that are self-regulated, consistent with the chemical enrichment of their previous generations of stars.
We thus construct a sub-sample considering only those SFGs that have $\rm\Delta_{MZR}\leq 0.25$. This results in a MZR selected sample of 2370 SFGs over the entire range of log(sSFR$\rm_{FIB}$) values, i.e., SFGs within the grey dashed lines in every panel of Figure~\ref{Fig:mmr}. To distinctly study the impact of galaxy mass on the O/Ar vs Ar plane (discussed in Section~\ref{sec:analysis}), we further separate the MZR selected sample into higher ($\geq9$~M$\rm_{\odot}$) and lower mass ($\leq8$~M$\rm_{\odot}$) sub-samples. The sample sizes are noted in Table~\ref{tab:sample}. The higher and lower mass samples have mean total stellar mass (± 1$\sigma$) of <log(M$\rm_{*}/$M$_{\odot}$)>$=9.41\pm0.29$ and $7.23\pm0.61$ respectively.


\begin{figure*}
\centering
\includegraphics[width=\textwidth]{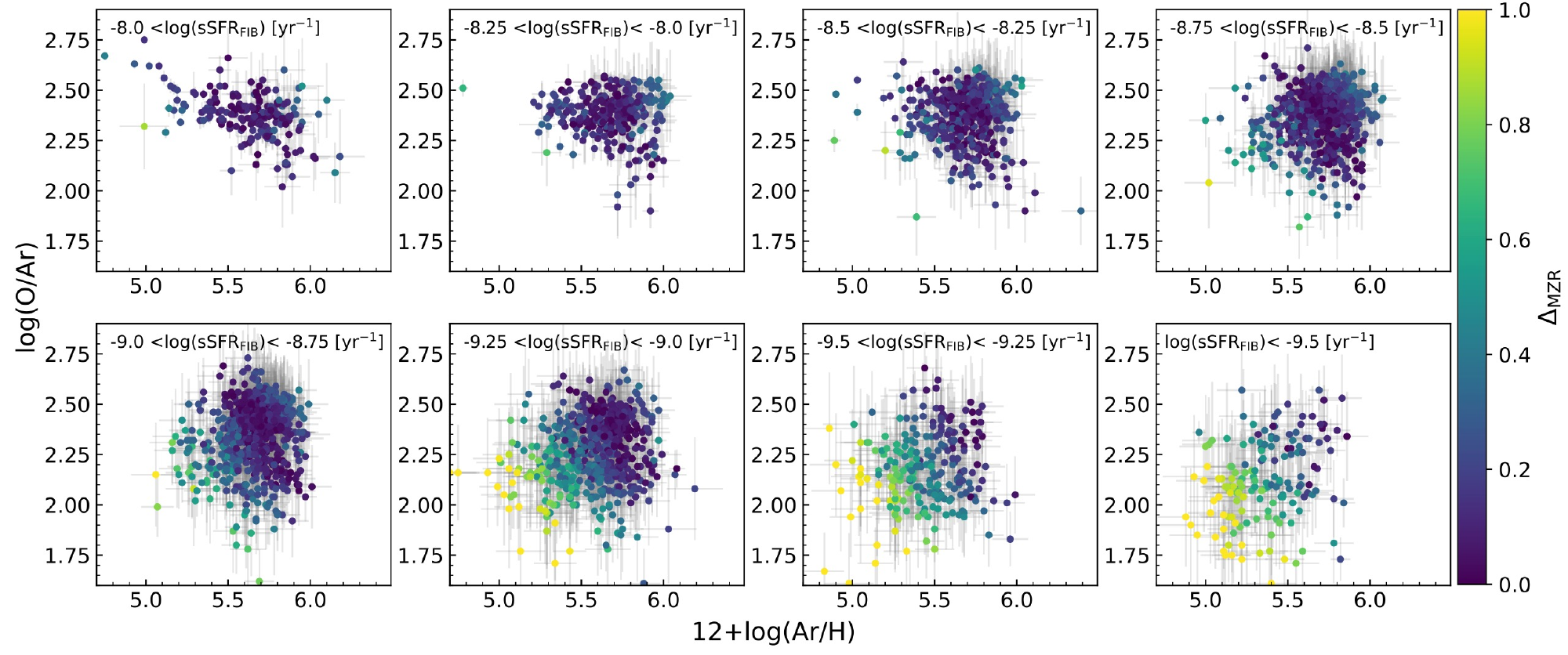}
\caption{Same as Figure~\ref{Fig:oar_obs} but now the SFGs in each panel are coloured by their $\rm\Delta_{MZR}$, i.e., the perpendicular distance from the empirical best-fit MZR shown in Figure~\ref{Fig:mmr} (see Section~\ref{sec:oar} for details). } 
\label{Fig:oar_delta}
\end{figure*}

\section{Chemical enrichment sequences of star-forming galaxies} 
\label{sec:analysis}

\begin{figure*}
\centering
\includegraphics[width=0.95\textwidth]{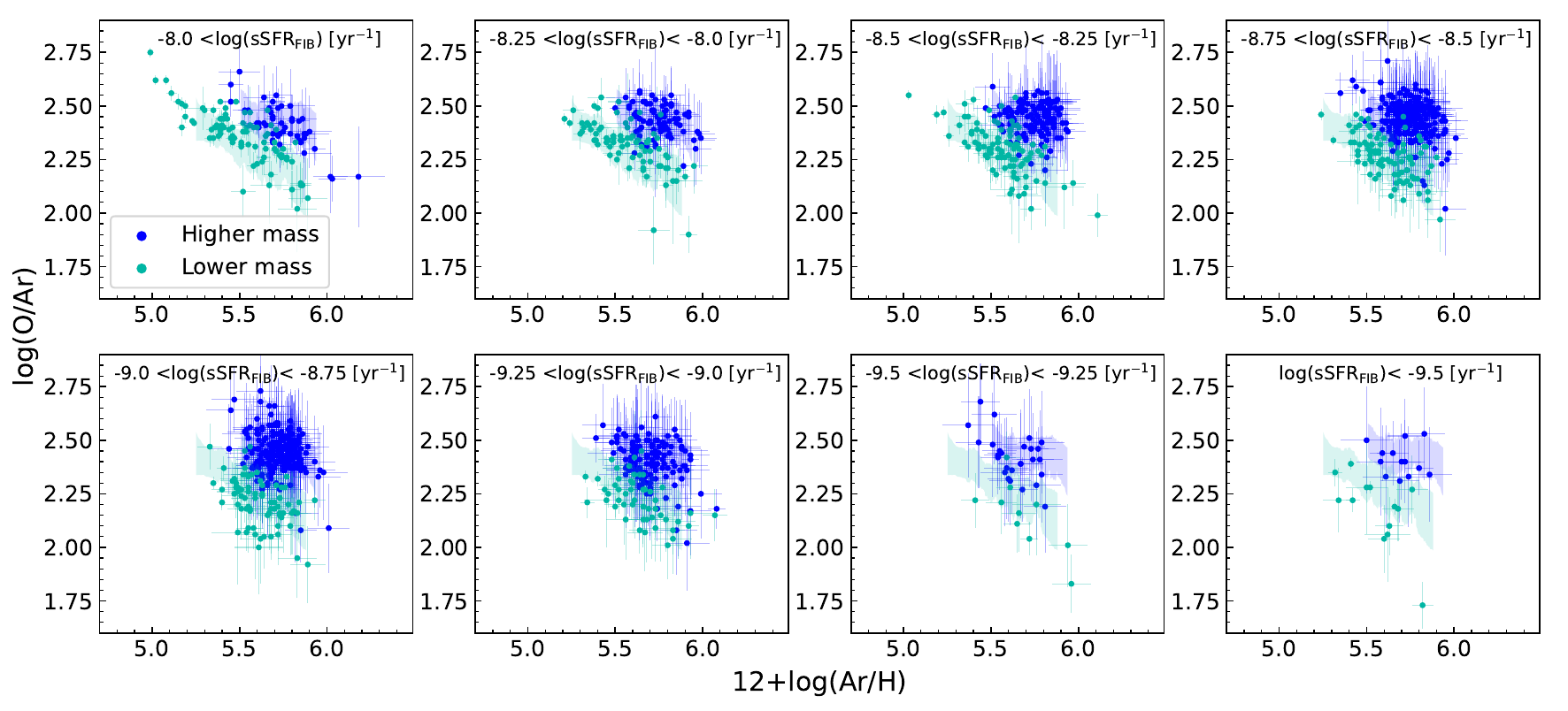}
\caption{Same as Figure~\ref{Fig:oar_obs} but now only showing the SFGs in the higher and lower mass MZR selected sub-samples (see Section~\ref{sec:mzr}). The standard deviations of the running mean of log(O/Ar) as a function of 12+log(Ar/H) over the entire range of log(sSFR$\rm_{FIB}$) is computed separately for both mass samples and shown with the shaded regions  (see Section~\ref{sec:seq} for details). } 
\label{Fig:oar_clean}
\end{figure*}


\subsection{The log(O/Ar) vs 12+log(Ar/H) plane of SFGs} 
\label{sec:oar}

Figure~\ref{Fig:oar_obs} shows the distribution of the abundance sample SFGs in the log(O/Ar) vs 12+log(Ar/H) plane, colored by their log(M$\rm_{*}$), separated in bins of log(sSFR$\rm_{FIB}$). The log(sSFR$\rm_{FIB}$) bin size has been chosen to ensure a wide range of Ar abundances (at least 1~dex) are spanned by SFGs in each bin, while allowing the maximum number of bins to check for any potential trend with log(sSFR$\rm_{FIB}$).  The panel corresponding to the highest log(sSFR$\rm_{FIB}$) bin in Figure~\ref{Fig:oar_obs} shows the distribution for the most complete ($\sim80$\%) sample of SFGs. Here we can clearly see that at fixed Ar abundance, higher mass galaxies exhibit higher log(O/Ar) values. Additionally, separate trends are seen for lower and higher mass galaxies with declining log(O/Ar) values with increasing Ar abundances. Subsequent panels in Figure~\ref{Fig:oar_obs} with decreasing log(sSFR$\rm_{FIB}$) are increasingly affected by incomplete sample selection (see discussion in Section~\ref{sec:final}), and the mass separation in the log(O/Ar) vs 12+log(Ar/H) plane is increasingly blurred.

In Figure~\ref{Fig:oar_delta}, we show the same SFGs in each panel but now coloured by their $\rm\Delta_{MZR}$. For the SFGs with the lowest $\rm\Delta_{MZR}$ values, there is a clear trend of decreasing log(O/Ar) values with increasing Ar abundances which is visible in all different log(sSFR$\rm_{FIB}$) bins. SFGs with larger $\rm\Delta_{MZR}$ values start appearing in the lower log(sSFR$\rm_{FIB}$) bins. SFGs that were offset from the empirical MZR line (Figure~\ref{Fig:mmr}) and exhibited low O abundances for their given masses, also exhibited low log(O/Ar) and low Ar abundance values, and blurred the mass dependent trends in log(O/Ar) vs 12+log(Ar/H) for the lower log(sSFR$\rm_{FIB}$) bins. The SFGs with $\rm\Delta_{MZR}>0.25$ are then removed and only the MZR selected sample is kept. The removed SFGs and the connection between the MZR-selection and the log(O/Ar) vs 12+log(Ar/H) plane is discussed further in the next Section.

Figure~\ref{Fig:oar_clean} shows the distribution of the higher and lower mass MZR sample SFGs in the log(O/Ar) vs 12+log(Ar/H) plane, separated in bins of log(sSFR$\rm_{FIB}$). The highest log(sSFR$\rm_{FIB}$) bin still shows the clear separation of higher and lower mass SFGs in the log(O/Ar) vs 12+log(Ar/H) plane. However, this mass separation is now clearly visible in all panels of Figure~\ref{Fig:oar_clean}. Removal of the $\rm\Delta_{MZR}>0.25$ SFGs allowed us to see the clear mass separation in the log(O/Ar) vs 12+log(Ar/H) plane, regardless of log(sSFR$\rm_{FIB}$). The separation in log(sSFR$\rm_{FIB}$) bins allows us to gauge the impact of sample selection (see discussion in Section~\ref{sec:final}) when sampling only those SFGs with direct abundances determined. Considering only those SFGs that follow the MZR and are thus self-regulated  (i.e., whose chemical evolution of previous generations of stars are imprinted on the ISM probed with direct abundance in the present star-forming episode), we find that regardless of their log(sSFR$\rm_{FIB}$) value, distinct sequences for higher and lower mass SFGs are seen in Figure~\ref{Fig:oar_clean}.



\subsection{Sequences in the log(O/Ar) vs 12+log(Ar/H) plane of SFGs} 
\label{sec:seq}

\begin{figure}
\centering
\includegraphics[width=\columnwidth]{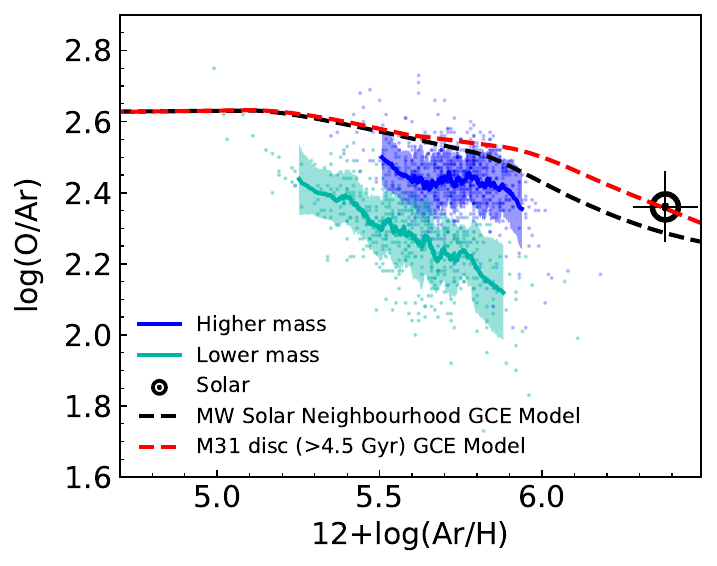}
\caption{Distribution of SFGs in the higher and lower mass MZR selected subsamples in the log(O/Ar) vs 12+log(Ar/H) plane. The running mean log(O/Ar) as a function of 12+log(Ar/H) is shown separately for higher and lower mass subsamples, with standard deviations shaded. Self-regulated GCE model for the MW solar neighbourhood and the self-regulated phase of our M31 disc model are marked. The solar value (\citealt{kob20sr}) is also marked.} 
\label{Fig:oar_final}
\end{figure}

Figure~\ref{Fig:oar_final} shows all the SFGs in the higher and lower mass MZR selected samples over all log(sSFR$\rm_{FIB}$) bins. We can see that at any given Ar abundance, higher mass SFGs clearly exhibit higher log(O/Ar) than lower mass SFGs. We compute a running mean (and standard deviation), separately for higher and lower mass SFGs, for the log(O/Ar) values as a function of Ar abundance. Figure~\ref{Fig:oar_final} thus shows the distinct sequences traced by higher and lower mass SFGs in the log(O/Ar) vs 12+log(Ar/H) plane\footnote{We note that the sequences in the log(O/Ar) vs 12+log(Ar/H) plane do not arise from any possible systematic errors in the ICF. See Appendix C in \citet{Bhattacharya25}.}. 

These median sequences are marked in Figure~\ref{Fig:oar_clean}. As the observed trend for the SFGs in the highest log(sSFR$\rm_{FIB}$) bin is nearly identical to that seen in Figure~\ref{Fig:oar_obs} prior to MZR selection, the consistency of the median trends with the clear trend seen for the SFGs in this bin confirms that the sequences are a consequence of mass-dependent chemical enrichment mechanisms and not simply a selection effect due to MZR selection. The median sequences marked in Figure~\ref{Fig:oar_clean} show that even if the higher and lower mass SFGs are separated by log(sSFR$\rm_{FIB}$), the individual SFGs are consistent with the same mass dependent trends across all log(sSFR$\rm_{FIB}$) bins. 

Figure~\ref{Fig:oar_final} also shows the self-regulated GCE models for the MW solar neighbourhood \citep{kob20sr} and the self-regulated phase (older than 4.5~Gyr, prior to the secondary gas infall; see \citealt{Kobayashi23}) for our M31 disc model. These are shown to illustrate the governing chemical enrichment sequence for galaxies having higher stellar mass than those in our SDSS sample. We can see the effect of galaxy mass on the chemical enrichment sequence in the log(O/Ar) vs 12+log(Ar/H) plane. The model for the highest mass galaxy, M31 (M$\rm_{*, disc+bulge}=$~1-1.5 $\times 10^{11}$M$_{\odot}$; \citealt{Tamm12}; see also \citealt{Bhattacharya23b} for a review of the mass determination history of M31), constrained from O and Ar abundances of $>4.5$~Gyr old M31 disc planetary nebulae,  shows the highest log(O/Ar) at any 12+log(Ar/H) value. This is followed closely by the MW (M$\rm_{*, disc+bar+bulge}=6.08 \pm 1.14 \times 10^{10}$M$_{\odot}$; \citealt{Licquia15}) whose solar neighbourhood GCE model, constrained by Fe and $\alpha$-element abundances from stars, traces a lower log(O/Ar) value at 12+log(Ar/H)$>5.7$. The higher mass SDSS galaxies with <M$\rm_{*}$>$\sim 2.6 \times 10^9$M$_{\odot}$ trace a lower log(O/Ar) sequence, while the lower mass SDSS galaxies with <M$\rm_{*}$>$\sim 1.7 \times 10^7$M$_{\odot}$ trace an even lower log(O/Ar) sequence.

The mass-dependent SFG sequences are blurred (Figure~\ref{Fig:oar_obs}) when considering galaxies that are offset from the empirical MZR for the less bursty SFGs (Figure~\ref{Fig:mmr}) where our abundance sample is less complete. It is likely that such SFGs have experienced outflow of enriched gas (e.g. \citealt{vanZee98}) or inflow of metal-poor gas (e.g. \citealt{Koppen05}) prior to star-formation, and are hence not self-regulated. We have shown for M31 (albeit using planetary nebulae; \citealt{Arnaboldi22,Kobayashi23}) that metal-poor gas infall can dilute the ISM leading to planetary nebulae with lower Ar and lower O/Ar values than the general trend (before subsequent extended star-formation brings the ISM back to the general self-regulated trend). It is thus the same ISM dilution that offsets the location of a SFG in the MZR plane, that also offsets its location from the general trend in the log(O/Ar) vs 12+log(Ar/H) plane. We note that a very small number of SFGs lie above the MZR in Figure~\ref{Fig:mmr} and appear to have higher log(O/Ar) than the general trend in Figure~\ref{Fig:oar_delta}. These may have had metal-rich gas infall. We exclude them in this work although their numbers are so small that their exclusion has negligible impact on the results. The ensemble of SFGs that follow the empirical MZR (as described in Section~\ref{sec:mzr}) likely have self-regulated chemical enrichment, wherein enrichment from star formation balances any potential gas infall, and traces mass-dependent loci in Figure~\ref{Fig:oar_final} in the log(O/Ar) vs 12+log(Ar/H) place.


\section{Discussion and Conclusion} 
\label{sec:discussion}

In this work, we use the log(O/Ar) vs 12 + log(Ar/H) plane for SFGs, analogous to the [$\alpha$/Fe] vs [Fe/H] plane for stars, to infer the mechanisms that govern galaxy chemical enrichment out to z$\sim$0.3. We obtain O \& Ar abundances for a sample of 3306 SFGs from the SDSS (see Section~\ref{sec:abund}). We then select those galaxies that emprically follow the MZR traced by the burstiest SFGs in our sample (and are thus forming the highest fraction of their stellar mass in the observed star-forming episode in a self-regulated manner), where we are most complete in obtaining abundances (see Sections~\ref{sec:final}~\&~\ref{sec:mzr}). We further restrict ourselves to higher (<M$\rm_{*}$>$\sim 2.6 \times 10^9$M$_{\odot}$) and lower mass (<M$\rm_{*}$>$\sim 1.7 \times 10^7$M$_{\odot}$) samples respectively. The SFGs with different mass bins clearly trace out distinct sequences in the log(O/Ar) vs 12 + log(Ar/H) plane, both in different sSFR bins (Figure~\ref{Fig:oar_clean}) and when considered altogether (Figure~\ref{Fig:oar_final}).

This is the first instance of such chemical enrichment sequences being presented in an ensemble of SFGs, here out to z$\sim$0.3. When considering self-regulated SFGs, we find no dependence on the chemical enrichment sequences on sSFR, while uncovering their dependence on stellar mass. The results are qualitatively consistent with the mass dependence of chemical enrichment sequences observed for the stars in the MW and its satellite galaxies \citep{Bensby14,Pompeia08,Tolstoy09,Kirby11,kob23book}, as well as the $\rm\sigma$ dependence of such sequences observed for nearby relatively massive ETGs \citep{Sybilska18}, and also absorption lines width (used as proxy for dynamical mass) dependence of such sequences observed for Damped Lyman-$\rm\alpha$ absorbers \citep[][]{Velichko24}. Additionally, the results are qualitatively consistent with the mass dependence of [$\alpha$/Fe] and [Fe/H] values (determined from stellar population model-fitting) of old self-regulated ETGs at z$\sim$0.05--0.06 \citep[][]{Thomas10}. 

The observed mass-dependent chemical enrichment sequences in the log(O/Ar) vs 12 + log(Ar/H) plane for SFGs are further qualitatively consistent with expectations from cosmological chemodynamical simulations by \citet{Vincenzo18}, and thereby allow for an interpretation of the observed results. Higher mass galaxies, that are more efficient at star-formation, produce more massive stars relatively rapidly, that in-turn rapidly produce more CCSNe which enrich the ISM to higher metallicities, here 12+ log(Ar/H), keeping relatively constant $\alpha$-abundances, here log(O/Ar). Then SNe Ia start erupting and subsequent generations of stars are formed from ISM with decreasing $\alpha$-abundances as metallicity, here 12+ log(Ar/H), is further increased. Lower mass galaxies, that are less efficient at star-formation, produce massive stars relatively slowly, that in-turn limits the enrichment of the ISM with CCSNe to lower metallicities allowing SNe Ia to start erupting when the ISM has lower metallicity and thus the decrease in $\alpha$-abundances sets-in at these lower metallicities. Such a mass-dependent chemical enrichment mechanism had previously been invoked to explain the observed lower [$\alpha$/Fe] at given [Fe/H] for the sequences traces by stars in the Magellanic clouds and other dwarf MW satellite galaxies  when compared with sequences traced by stars in the more massive MW. 
We thus find that differences in star-formation efficiency in galaxies leads to differences in supernova enrichment, deciphered either from abundances of stars within a galaxy (as illustrated in Figure~\ref{Fig:oar_final} for the MW \& M31) or from an ensemble of SFGs as shown in this work. This results in mass-dependent chemical enrichment sequences for SFGs in the log(O/Ar) vs 12 + log(Ar/H) plane.

Given the direct O \& Ar abundances determined for 11 SFGs at z$\sim$1.3--7.7 observed with JWST/NIRSPEC and Keck/MOSFIRE \citep{Bhattacharya25}, MW-like self-regulated chemical enrichment sequences and their underlying CCSNe and SNe Ia interplay dominated mechanisms, shown here to be prevalent in ensembles of SFGs out to z$\sim$0.3, may be in place as early as z$\sim$1.3--4 (see also Figure~4 [Left] in \citealt{Bhattacharya25}, also showing the higher and lower mass sequences shown in Figure~\ref{Fig:oar_final}). The same mechanisms may be at play at even higher redshift galaxies (z$\sim$3.3--7.7) but in conjunction with intermittent star-formation, although additional non-standard sources of chemical enrichment may also have influence at such redshifts \citep{Bhattacharya25}. 

While the SDSS survey is limited to relatively metal-poor (significantly sub-solar; see Figure~\ref{Fig:oar_final}) and relatively bursty and lower mass SFGs for such an analysis, upcoming large ground-based spectroscopic surveys \citep[e.g. the Prime Focus Spectrograph galaxy evolution survey at Subaru;][]{pfs} should make it possible to build-up a large sample of SFGs, covering a wider parameter space of metallicity, mass and sSFR, with direct determinations of O and Ar abundances from auroral line-flux measurements.  The log(O/Ar) vs 12 + log(Ar/H) plane thus offers a new diagnostic window for constraining galaxy chemical enrichment mechanisms for a wide array of SFGs with diverse properties, covering a wide range of redshifts.


\section*{Acknowledgements}

We thank the anonymous referee for their valuable suggestions. SB was supported by the INSPIRE Faculty award (DST/INSPIRE/04/2020/002224), Department of Science and Technology (DST), Government of India. SB and MA acknowledge support to this research from the European Southern Observatory, Garching, through the 2022 SSDF. SB and OG acknowledge support to this research from Excellence Cluster ORIGINS, which is funded by the Deutsche Forschungsgemenschaft (DGF, German Research Foundation) under Germany's Excellence Strategy - EXC-2094-390783311. MA and OG thank the Research School of Astronomy and Astrophysics at ANU for support through their Distinguished Visitor Program in 2024. This work was supported by the DAAD under the Australia-Germany joint research programme with funds from the Australian Ministry for Science and Education. CK acknowledges funding from the UK Science and Technology Facility Council through grant ST/Y001443/1.

\section*{Data Availability}

Based on tabulated data publicly available in the \href{https://www.sdss4.org/dr17/spectro/galaxy_mpajhu}{MPA-JHU SDSS catalogue}. GCE models can be shared upon reasonable request. 



\bibliographystyle{mnras}
\bibliography{oar} 

\appendix

\section{Impact of utilising fiber vs. total specific star formation rate}
\label{app:fib_tot}

\begin{figure}
\centering
\includegraphics[width=0.88\columnwidth]{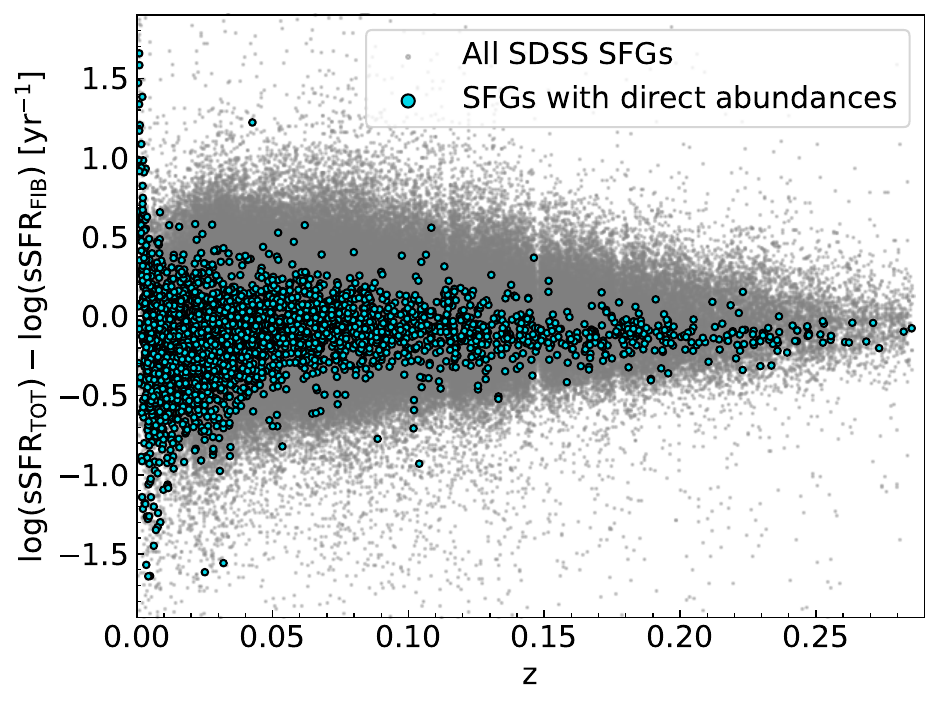}
\caption{The difference between log(sSFR$\rm_{TOT}$) from log(sSFR$\rm_{FIB}$) as a function of redshift for all SFGs out to z$\sim$0.3 from the MPA-JHU SDSS catalogue (grey) and those with direct O and Ar abundances determined (cyan).} 
\label{Fig:fib_tot}
\end{figure}

Depending on the observed size of a galaxy, the 3$''$ diameter SDSS fiber placed at the center of each galaxy may not cover its entire luminous body. Given the chemical inhomogeneity of SFGs \citep[e.g.][]{Maiolino19}, the O \& Ar abundances would only reflect the stellar population properties of the region spanned by the SDSS fiber in such a scenario. We thus utilize the determined sSFR$\rm_{FIB}$ value (available as part of the MPA-JHU catalog; computed from both SFR and stellar mass estimated within the fiber diameter) for our analysis rather than the sSFR$\rm_{TOT}$ value that is averaged over the entire galaxy. Furthermore, as shown in Figure~\ref{Fig:fib_tot}, only a small number of very low-z SFGs with direct O \& Ar abundances have large difference between log(sSFR$\rm_{FIB}$) and log(sSFR$\rm_{TOT}$),  up to $-1.7$ [yr$^{-1}$], with a vast majority of these SFGs having a difference $<0.3$ [yr$^{-1}$]. This difference is lower for our sample of SFGs with direct abundances than for the full sample of SFGs surveyed by SDSS (Figure~\ref{Fig:fib_tot}), a potential consequence of our selection of lower mass SFGs. 

For our analysis, we can conservatively consider only those SFGs that have log(sSFR$\rm_{TOT}$) $-$ log(sSFR$\rm_{FIB}$) between $-0.3$ \& $0.3$ [yr$^{-1}$]. Such a selection will result in 2480 SFGs from our sample of 3306 SFGs with direct abundances determined, excluding primarily lower mass low  log(sSFR$\rm_{FIB}$) SFGs. We found that even with such a conservative selection that excludes $\sim25$\% of our sample, the mass dependent trends in the O/Ar vs Ar abundance plane are clearly seen for all sSFR bins (as in Figure~\ref{Fig:oar_clean}) but with fewer SFGs (especially lower mass ones) present in the lower sSFR bins. Thus, the analysis and results of this work remain valid even if sSFR$\rm_{TOT}$ is used instead of sSFR$\rm_{FIB}$, or if only SFGs with low log(sSFR$\rm_{TOT}$) $-$ log(sSFR$\rm_{FIB}$) are conservatively selected.

\section{Star-forming main sequence}
\label{app:SFMS}

\begin{figure*}
\centering
\includegraphics[width=\textwidth]{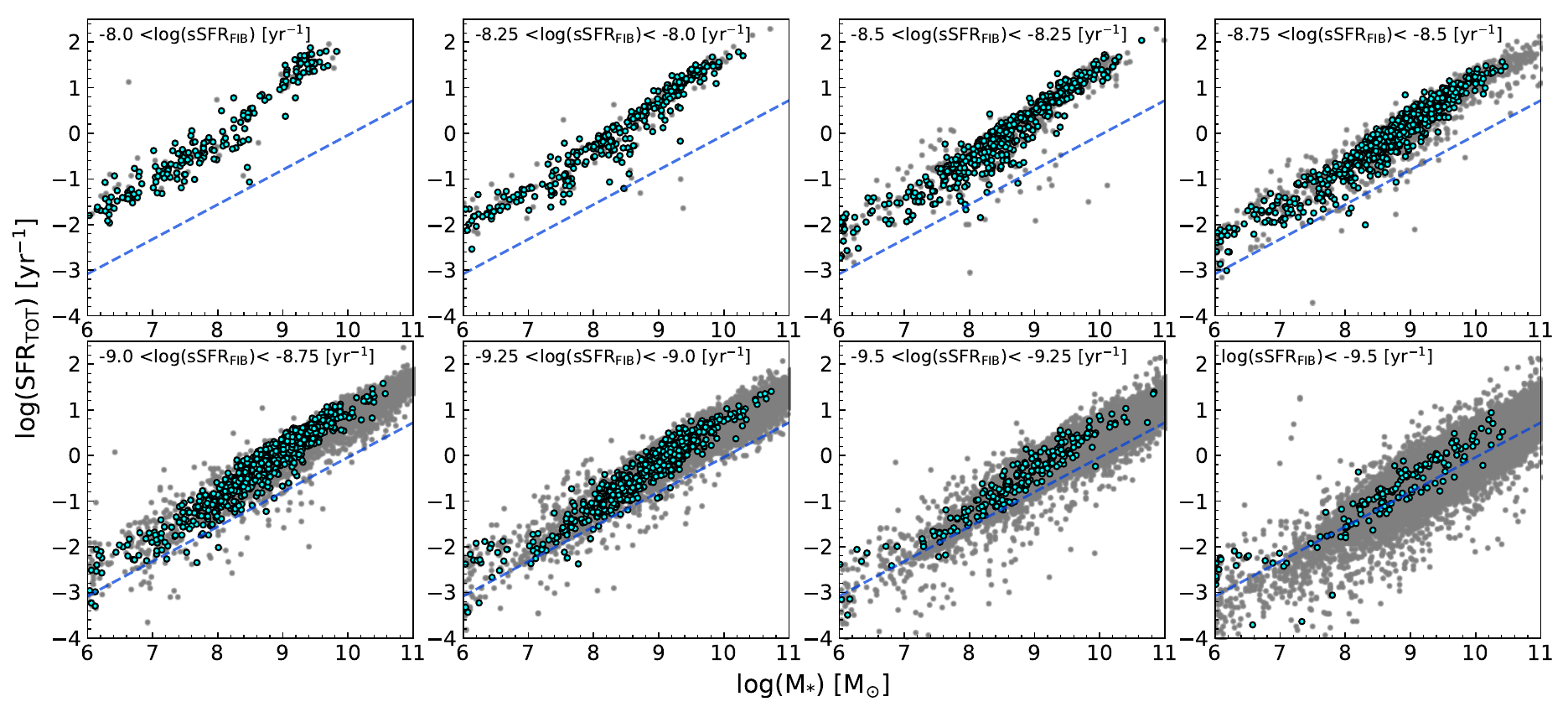}
\caption{Stellar mass vs. total SFR for all SFGs out to z$\sim$0.3 from the MPA-JHU SDSS catalogue (grey) and those with direct determination of O \& Ar abundances (cyan), separated in bins of log(sSFR$\rm_{FIB}$). The dashed blue line marks the objective definition of the main sequence of SFGs by \citet{Renzini15}. } 
\label{Fig:sf_ms}
\end{figure*}

The star-forming main sequence (SFMS) of galaxies refers to the tight correlation between stellar mass and log(SFR$\rm_{TOT}$) observed for SFGs (e.g. \citealt{Noeske07,Popesso19}). As the exact correlation depends on the selection of SFGs that may vary for different surveys, \citet{Renzini15} computed an objective equation to define the SFMS from SDSS DR7 SFGs to faciliate comparison between different survey results. Figure~\ref{Fig:sf_ms} shows the distribution of all SFGs out to z$\sim$0.3 from the MPA-JHU SDSS catalogue, as well as the SFGs with direct O \& Ar abundances in the different log(sSFR$\rm_{FIB}$) bins utilised in this work, along with the objective SFMS from \citet{Renzini15}. Its clear that for the highest  log(sSFR$\rm_{FIB}$) bin, SFGs are more strongly star-forming than the objective SFMS at any given mass. As we go to lower and lower log(sSFR$\rm_{FIB}$) bins, the SFGs distribute closer and closer to the objective SFMS. For all log(sSFR$\rm_{FIB}$) bins, the distribution of SFGs with direct O \& Ar abundances in the log(M$_*$) vs. log(SFR$\rm_{TOT}$) plane are consistent with that of the full sample. 


\bsp	
\label{lastpage}
\end{document}